# Accounting for iron-related off-target binding effects of $^{18}$F-AV1451 PET in the evaluation of cognition and microstructure in APOE-ε4+ MCI

Jason Langley[1], Daniel E. Huddleston[2], Ilana J. Bennett[3], and Xiaoping P. Hu[1,4] for the Alzheimer's Disease Neuroimaging Initiative

[1] Center for Advanced Neuroimaging, University of California Riverside, Riverside, CA, USA
[2] Department of Neurology, Emory University, Atlanta, GA, USA
[3] Department of Psychology, University of California Riverside, Riverside, CA, USA
[4] Department of Bioengineering, University of California Riverside, Riverside, CA, USA

E-mail: xhu ‚at' engr.ucr.edu

## Abstract

The pathology of Alzheimer's disease (AD) and mild cognitive impairment (MCI) is characterized by the presence of β-amyloid (Aβ) extracellular plaques and neurofibrillary tangles containing hyper-phosphorylated tau. Individuals carrying the apolipoprotein E-ε4 (APOE-ε4) allele are at increased risk of cognitive decline and developing AD pathology. The development of positron emission tomography (PET) radioligands sensitive to tau neurofibrillary tangles, such as $^{18}$F-AV1451, has allowed for visualization and assessment of AD pathology *in vivo*. However, the radioligand used in $^{18}$F-AV1451 binds with iron in addition to tau neurofibrillary tangles. Here, we employ multimodal neuroimaging analyses, combining iron-sensitive measures from MRI with $^{18}$F-AV1451 PET, to examine off-target binding effects of the $^{18}$F-AV1451 radioligand in cohorts of 20 APOE-ε4 negative, 20 APOE-ε4 positive MCI participants, and 29 control participants. Increased tau pathology ($^{18}$F-AV1451 PET uptake), after controlling for tissue susceptibility, was found in the temporal lobe and hippocampus of APOE-ε4+ MCI participants as compared to APOE-ε4 negative MCI and control participants. Tau pathology in the hippocampus was significantly related to memory, but only in APOE-ε4+ participants. Correlations between hippocampal $^{18}$F-AV1451 PET uptake and cognitive correlations did not significantly differ when correcting for the influence of iron on $^{18}$F-AV1451 PET signal. However, controlling for susceptibility was found to influence correlations between tau-PET uptake and diffusion metrics and the change in this interaction may be due to the influence of iron on diffusivity. Taken together, these results suggest that iron does not need to be accounted for in group comparisons of tau-PET uptake or correlations between cognitive measures and tau-PET SUVR. However, iron should be taken into account in correlations between diffusion measures and tau-PET uptake.

## 1. Introduction

Alzheimer's disease (AD) is the most frequent cause of dementia and it is estimated that 1 in 85 people worldwide will develop AD by 2050.[1] Patients with mild cognitive impairment (MCI) exhibit declines in cognitive performance that do not meet the threshold for dementia, but are likely to later convert to AD.[2] AD and MCI pathology is characterized by the presence of β-amyloid (Aβ) extracellular plaques and neurofibrillary tangles containing hyper-phosphorylated tau.[3-5] Both Aβ plaque and tau neurofibrillary tangle deposition precede cognitive decline,[6,7] However, in contrast to Aβ pathology, which is weakly associated with cognitive decline,[8] tau pathology is strongly associated with neurodegeneration and cognitive impairment.[9]

Individuals carrying the apolipoprotein E-ε4 (APOE-ε4) allele are at increased risk of cognitive decline and developing AD pathology.[10] The APOE-ε4 allele is associated with more severe cognitive impairment in individuals with AD or MCI[11-13] as well as more rapid age-related cognitive decline in healthy older adults.[14] These cognitive effects are thought to reflect AD pathology. Both animal[15,16] and human stem cell[17] models expressing the APOE-ε4 allele have shown increased tau pathology. Magnetic resonance imaging (MRI) studies have similarly found atrophy in the hippocampus and temporal lobe of healthy older adults as well as cognitively impaired subjects with the APOE-ε4 allele.[18]

The development of positron emission tomography (PET) radioligands sensitive to tau neurofibrillary tangles, such as $^{18}$F-AV1451, has allowed for visualization and assessment of AD pathology *in vivo*. Increases in tau-PET signal have been reported in individuals with MCI or AD, regardless of APOE-ε4 status, relative to controls.[19-24] Imaging studies using $^{18}$F-AV1451 have found tau-PET signal to be associated with lower brain volume[25] and cortical thickness[26-28] derived from T$_1$-weighted images. Other studies examining tau-PET signal and microstructural measures from diffusion tensor imaging, a MRI contrast sensitive to the motion of water molecules,[29] have found tau-PET signal is related to microstructure measures associated with neurodegeneration[30-32] in AD.

However, the radioligand used in $^{18}$F-AV1451 binds with iron in addition to tau neurofibrillary tangles,[33-36] which may confound these associations. In particular, strong associations in the cortex have been observed when cortical tau-PET uptake is compared to tissue susceptibility,[37,38] a MRI measure sensitive to iron.[39] Elevated cortical iron levels have been observed in AD and MCI[40-43] and off target binding of the tau radioligand may confound interpretation of cross-sectional or longitudinal changes in $^{18}$F-AV1451 images, because changes in tau-PET signal could be due to iron or tau.

Multimodal neuroimaging analyses, combining iron-sensitive measures from MRI with $^{18}$F-AV1451 PET, allow for removal of off-target binding effects of the $^{18}$F-AV1451 radioligand. Adjustments to address off-target binding are important given that individuals with one or more APOE-ε4 allele (APOE-ε4 positive) have higher cortical iron burden.[40,43,44] Here, we use tissue susceptibility to control for iron-related off-target binding effects in $^{18}$F-AV1451 PET images and explore the relationship between tau-deposition, tissue microstructure, cognition, and APOE-ε4 carrier status in MCI. Demonstrating that the previously reported relationships between tau-PET signal and neurodegeneration or cognitive impairment remain significant in the hippocampus and temporal lobe after controlling for susceptibility would provide confidence that they are in fact due to tau and not iron.

## 2. Methods

### 2.1 ADNI Overview

Data used in this study were obtained from the ADNI database (adni.loni.usc.edu). ADNI was launched in 2003 as a public-private partnership supported project. The ADNI was launched in 2003 as a public-private partnership, led by Principal Investigator Michael W. Weiner, MD. The primary goal of ADNI is to test whether serial MRI, PET, other biological markers, and clinical and neuropsychological assessments can be combined to measure the progression of MCI and early AD. Up-to-date information can be found at www.adni-info.org. The ADNI data were collected from over 50 research sites and the ADNI study was approved by the local Institutional Review Boards (IRBs) of all participating sites. The detailed information and complete list of ADNI sites' IRBs could be found at http://adni.loni.usc.edu/about/centers-cores/study-sites/ and http://www.adni-info.org/.

Study subjects and, if applicable, their legal representatives, gave written informed consent at the time of enrollment for imaging data, genetic sample collection and clinical questionnaires. Exclusion criteria determined by ADNI was followed. Subjects were excluded from the analysis if they had Parkinson's disease, Huntington's disease, progressive supranuclear palsy, a history of seizures, normal pressure hydrocephalus, brain tumors, multiple sclerosis, subdural hematoma, a history of head trauma, known brain structural abnormalities, a history of major depression, schizophrenia, alcohol or substance abuse, bipolar disorder, or currently using psychoactive medications. Individuals with contraindications to MRI imaging such as pacemakers, heart valves, or other foreign objects or implants in the body were excluded.

### 2.2 Participants

The ADNI3 database was queried for individuals with tau-sensitive PET ($^{18}$F-AV1451), multi-shell diffusion-weighted images, and multi-echo gradient echo MRI images at the same scanning visit, as well as APOE-ε4 status. From this cohort, we selected all individuals with a diagnostic status of MCI or control at the time of the visit, which included 20 APOE-ε4 negative (ε4-) and 20 APOE-ε4 positive (ε4+) MCI participants and 29 APOE-ε4- control participants. Imaging data were downloaded in December 2019.

All MCI participants in the ADNI3 database had a subjective memory concern reported by a clinician, abnormal memory function on the education-adjusted Logical Memory II subscale, and a clinical dementia rating greater than 0.5. Further, MCI participants were deemed to have cognitive and functional performance that was sufficiently intact to not merit a diagnosis of attention deficit disorder by the site physician.

### 2.3 MRI Acquisition

All MRI data used in this study were acquired on Siemens Prisma or Prisma fit scanners. Anatomic images were acquired with an MP-RAGE sequence (echo time (TE)/repetition time (TR)/inversion time=2.98/2300/900 ms, flip angle=9°, voxel size=1.0×1.0×1.0 mm$^3$, and GRAPPA acceleration factor=2) and were used for registration to common space and correction of partial volume effects in the PET data.

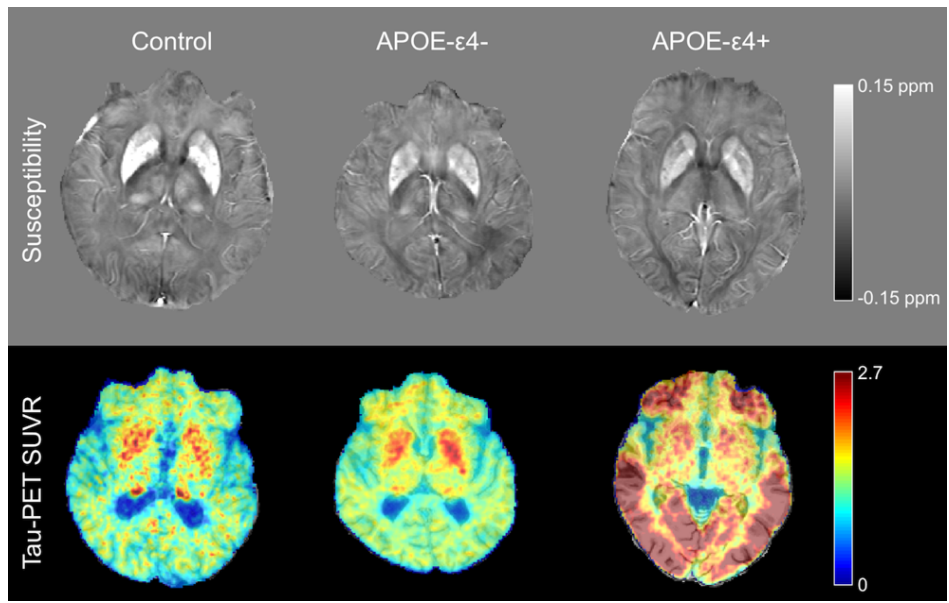

**Figure 1.** Illustrations of typical susceptibility (top row) and tau-PET SUVR (bottom row) images in subjects from the control (left column), APOE-ε4- MCI (middle column), and APOE-ε4+ MCI (right column) groups.

Iron-sensitive data were collected with a three-echo 2D gradient recalled echo (GRE) sequence (TE$_1$/ΔTE/TR = 6/7/650 ms, flip angle=20°, field of view=220×220 mm$^2$, matrix size of 256×256, 44 slices, slice thickness=4.0 mm) and used for measurement of brain iron.

Diffusion weighted imaging (DWI) data were acquired with a multiband diffusion weighted echo planar imaging (EPI) spin echo sequence (TE/TR=71/3400ms, field of view = 232×232 mm$^2$, voxel size=2×2×2mm$^3$, multiband acceleration factor=3, PA phase encoding direction). Diffusion weighting was applied in 54 directions with b values of 1000 and 2000 s/mm$^2$. A two-echo 2D GRE sequence (TE$_1$/TE$_2$/TR=4.92/7.38/571 ms, flip angle=60°, voxel size=3.0×3.0×3.0 mm$^3$) was used for correction of susceptibility distortion in the diffusion images.

### 2.4 T$_1$-common space registration

Transforms for MRI imaging data were derived with FMRIB Software Library (FSL). A transformation was derived between individual subject space to 2 mm Montreal Neurological Institute (MNI) T$_1$-weighted space using FMRIB's Linear Image Registration Tool (FLIRT) and FMRIB's Nonlinear Image Registration Tool (FNIRT) in the FSL software package using the following steps.[45, 46] First, the T$_1$-weighted image was skull stripped using the brain extraction tool (BET). Next brain extracted T$_1$-weighted images were aligned with the MNI brain extracted image using an affine transformation. Finally, a nonlinear transformation was used to generate a transformation from individual T$_1$-weighted images to T$_1$-weighted MNI common space.

### 2.5 QSM Processing

Susceptibility images were constructed using the following procedure. First, a brain mask was derived from the first echo of the magnitude data. Next, the brain mask was carefully examined and any areas of the mask outside the brain were manually removed. Background phase was removed using harmonic phase removal using the Laplacian operator (iHARPERELLA).[47] Finally, susceptibility maps were derived from the frequency map of brain tissue using an improved least-squares (iLSQR) method[48, 49] and Laplace filtering with a threshold of 0.04 as a truncation value. All susceptibility images were processed in MATLAB (The MathWorks, Inc., Natick, MA, USA) using STISUITE. The resulting susceptibility maps were aligned to each subject's T$_1$-weighted image using a rigid body transform derived via the magnitude image from the first echo.

### 2.6 DWI Processing

Diffusion data were preprocessed with FSL[45, 50, 51] and were first corrected for motion and eddy currents using EDDY. Next, field maps were constructed and used to correct magnetic field inhomogeneities in the diffusion images using FUGUE. Finally, the *b*=0 image was brain extracted and a transform between each subject's T$_1$-weighted and *b*=0 images was derived using a rigid body transform with a boundary-based registration cost function.

Single-compartment parameters (fractional anisotropy, FA; mean diffusivity, MD) were derived from the diffusion data using DTIFIT. Advanced modeling was performed using the NODDI toolbox v1.0.1 (http://www.nitrc.org/projects/noddi_toolbox) in MATLAB

Table 1. Demographic information for the groups used in this analysis. Data is presented as mean ± standard error. One-way analysis of variances (ANOVAs) were used for group comparisons of age, education, and cognition from which *p* values are shown.

|  | Control | MCI | | Group difference |
|---|---|---|---|---|
|  |  | APOE-ε4- | APOE-ε4+ |  |
| N (M/F) | 29 (13/16) | 20 (11/9) | 20 (13/7) |  |
| Age | 73.7±1.5 | 75.6±1.8 | 74.4±1.9 | 0.733 |
| Education | 16.3±0.5 | 15.6±0.6 | 15.4±0.6 | 0.417 |
| MOCA | 24.7±0.6 | 21.9±0.8 | 20.7±0.8 | $<10^{-4}$ |
| MMSE | 28.8±0.4 | 28.2±0.5 | 26.9±0.5 | 0.02 |
| CDR | 0.0±0.0 | 0.5±0.0 | 0.5±0.0 | $<10^{-4}$ |
| ADAS13 | 13.7±0.9 | 16.8±1.2 | 21.4±1.4 | $<10^{-4}$ |
| ADAS Delayed Recall | 3.1±0.3 | 4.4±0.5 | 6.3±0.5 | $<10^{-4}$ |

[52]. NODDI fitting was performed using the default settings and maps of cerebrospinal fluid (CSF) volume fraction (denoted fiso) and the fraction of water in the restricted compartment (ficvf) were generated.

### 2.7 PET Acquisition and Processing

The radiochemical synthesis of $^{18}$F-AV1451 was overseen and regulated by Avid Radiopharmaceuticals and distributed to the qualifying ADNI sites where PET imaging was performed according to standardized protocols. The $^{18}$F-AV-1451 protocol entailed the injection of 10 mCi of tracer followed by an uptake phase of 80 min during which the subjects remained out of the scanner, and then collection of the $^{18}$F-AV-1451 emission data as 4×5min frames. PET with computed tomography imaging (PET/CT) scans preceded these acquisitions with a CT scan for attenuation correction; PET-only scanners performed a transmission scan following the emission scan.

PET imaging data were analyzed with FSL and PET partial volume correction (PETPVC) toolbox [53]. Motion was corrected in $^{18}$F-AV1451 PET scans were co-registered to the first frame and averaged using rigid-body transforms with FLIRT in FSL. Next, the motion-corrected mean PET scans were registered to the participant's own T$_1$-weighted MRI image using a rigid-body transform with a normalized mutual information cost function in FLIRT. Grey matter, white matter, and CSF maps were segmented in the T$_1$-weighted MRI image and used to correct for partial volume effects using PETPVC.[53] A combination of Labbé[54] and region-based voxel-wise correction [55] was chosen to mitigate sensitivity to point spread function mismatch. The median standardized uptake value (SUV) in left + right cerebellar cortex was chosen as a reference. Figure 1 shows a comparison of typical susceptibility maps and SUV ratios (SUVR) for a subject from each group.

### 2.8 Regions of Interest

Atlases from the Harvard-Oxford subcortical atlas and a prior study parceling the cortex[56] were used to define standard space regions of interest (ROIs) in bilateral hippocampus and temporal lobe, respectively. The ROIs were then transformed from MNI space to subject space using linear and nonlinear transforms in FSL as described in the earlier sections.

Each aligned ROI was thresholded at 60% and binarized. To ensure that signal from white matter did not contaminate measures in the temporal lobe, the binarized temporal lobe ROI was multiplied by each individual's grey matter mask. Mean single-compartment diffusion (FA, MD), NODDI (fiso, ficvf), tau-PET SUVR, and susceptibility were measured in each resultant ROI for each participant.

### 2.9 Statistical Analysis

All statistical analyses were performed using IBM SPSS Statistics software version 24 (IBM Corporation, Somers, NY, USA) and results are reported as mean ± standard error. A *P* value of 0.05 was considered significant for all statistical tests performed in this work. Normality of tau-PET, diffusion, and iron data was assessed using the Shapiro-Wilk test for each group and all data was found to be normal.

The effect of group (APOE-ε4+ MCI, APOE-ε4- MCI, control) was tested with separate analysis of covariance (ANCOVA) in each ROI for tau-PET SUVR, susceptibility, and each single compartment diffusion index (FA, MD), controlling for sex and iron. Iron was controlled here since it is an off-target bind for the $^{18}$F-AV1451 radioligand and it correlates with single-compartment diffusion indices [57]. For all ANCOVAs, if the interaction was significant, post hoc comparisons between each pair of groups were performed using respective two-tailed t-tests. The effect of group was tested in each ROI for NODDI measure (fiso, ficvf) with separate ANCOVAs, controlling for sex. Iron-related off-target binding effects were assessed by examining the relationship between tau-PET SUVR and susceptibility

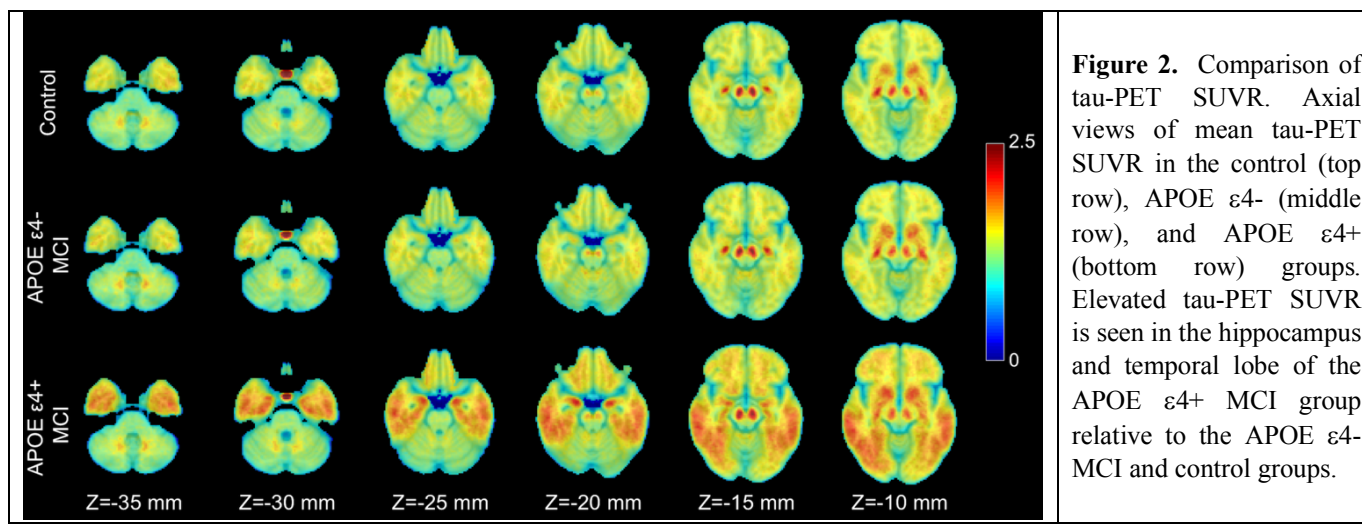

**Figure 2.** Comparison of tau-PET SUVR. Axial views of mean tau-PET SUVR in the control (top row), APOE ε4- (middle row), and APOE ε4+ (bottom row) groups. Elevated tau-PET SUVR is seen in the hippocampus and temporal lobe of the APOE ε4+ MCI group relative to the APOE ε4- MCI and control groups.

using Pearson correlations in each ROI, separately for each group.

The impact of iron on the relationship between tau-PET SUVR and both microstructural measures (FA, MD, fiso, and ficvf) and cognitive measures (delayed word recall and ADAS13) was assessed by performing separate multiple regressions that excluded susceptibility as a predictor in the second step. For each dependent measure, an $R^2$-change $F$-test was used to statistically compare the models with and without controlling for susceptibility.

## 3. Results

### 3.1 Sample Demographics

Demographic data for each group is shown in Table 1. Age ($P=0.234$; $F=1.479$) and education ($P=0.367$; $F=1.105$) exhibited no significant group effect. As expected, significant group effects were observed in MOCA ($P<10^{-4}$; $F=8.986$), MMSE ($P=0.022$; $F=4.078$), and CDR ($P<10^{-4}$; $F=186.746$) with both MCI groups showing reduced scores on MOCA (APOE-ε4+: $P=0.0002$; APOE-ε4-: $P=0.005$) and CDR (APOE-ε4+:$P<10^{-4}$; APOE-ε4-: $P<10^{-4}$) relative to the control group, whereas only the APOE-ε4+ MCI group had a lower MMSE score than the control group ($P=0.006$). Significant group effects were also seen in ADAS delayed recall ($P<10^{-4}$; $F=12.274$) and ADAS13 ($P<10^{-4}$; $F=16.453$). Higher ADAS delated recall scores were seen in the APOE-ε4+ MCI group relative to the APOE-ε4- MCI ($P=0.011$) and control ($P<10^{-4}$) groups, and in the APOE-ε4- MCI group relative to the control group ($P=0.031$). Higher ADAS13 scores were seen in the APOE-ε4+ MCI group relative to the APOE-ε4- MCI ($P=0.001$) and control ($P<10^{-4}$) groups, and in the APOE-ε4- MCI group relative to controls ($P=0.037$).

### 3.2 Effects of Group

The effect of group (APOE-ε4+ MCI, APOE-ε4- MCI, control) on susceptibility was assessed with separate ANCOVAs for each ROI (temporal lobe, hippocampus) with sex as a covariate. For the temporal lobe, a significant main effect of group ($P=0.040$; $F=3.401$) revealed higher susceptibility in the APOE-ε4+ MCI group (0.011 ppm ± 0.001 ppm) compared to APOE-ε4- MCI (0.007 ppm ± 0.001 ppm; $P=0.015$) and a similar trend compared to the control group (0.008 ppm ± 0.001 ppm; $P=0.057$), the latter of which did not differ from each other ($P=0.464$). For the hippocampus, a significant main effect of group ($P=0.013$; $F=4.712$) revealed higher susceptibility in the APOE-ε4+ MCI group (0.027 ppm ± 0.003 ppm) relative to the APOE-ε4- MCI (0.013 ppm ± 0.003 ppm, $P=0.004$) and control (0.017 ppm ± 0.003 ppm, $P=0.032$) groups, the latter of which did not differ from each other ($P=0.331$).

The effect of group on tau-PET SUVR was assessed with separate ANCOVAs for each ROI, with sex and ROI susceptibility as covariates (Figures 2-3). For the temporal lobe, a significant main effect of group ($P=3.27×10^{-4}$; $F=9.20$) revealed higher tau-PET SUVR in the APOE-ε4+ MCI group (1.366 ± 0.043) relative to the APOE-ε4- MCI (1.112 ± 0.046; $P=0.0002$) and control (1.164 ± 0.036; $P=0.001$) groups, the latter of which did not differ from each other ($P=0.382$). For the hippocampus, a significant main effect of group ($P=0.049$; $F=3.179$) similarly revealed higher tau-PET SUVR in the APOE-ε4+ MCI group (1.532 ± 0.047) relative to the APOE-ε4- MCI (1.362 ± 0.051; $P=0.022$) and control (1.409 ± 0.037; $P=0.045$) groups, the latter of which did not differ from each other ($P=0.300$).

The effect of group on microstructure was assessed with separate ANCOVAs for each ROI and diffusion parameter (FA, MD, fiso, ficvf), with sex as a covariate (Figure 3). For the temporal lobe, results revealed significant main effects of group for FA ($P=0.003$; $F=6.236$), MD ($P=0.009$; $F=5.060$), and fiso ($P=0.040$; $F=3.384$), but not ficvf ($P=0.337$;

$F$=1.108). Decreases in temporal lobe FA were seen in the APOE-ε4- (0.185 ± 0.002; *P*=0.038) and APOE-ε4+ (0.182 ± 0.002; *P*=0.001) MCI groups relative to the control group (0.191 ± 0.002). An increase was observed in MD of the APOE-ε4+ MCI group ($9.66\times10^{-4}$ ± $1.7\times10^{-5}$ mm$^2$/s) relative to the control group ($8.95\times10^{-4}$ ± $1.7\times10^{-5}$ mm$^2$/s; *P*=0.002). No difference in temporal lobe MD was seen between the APOE-ε4+ MCI and APOE-ε4- MCI groups (ε4-: $9.32\times10^{-4}$ ± $1.9\times10^{-5}$ mm$^2$/s; *P*=0.180) or APOE-ε4- and control groups (*P*=0.128). An increase in fiso was seen in the APOE-ε4+ MCI group (0.319 ± 0.013) relative to the control group (control: 0.277 ± 0.010; *P*=0.013). No difference in fiso was observed between the APOE-ε4+ MCI and APOE-ε4- MCI groups (ε4-: 0.302 ± 0.014; *P*=0.399) or between APOE-ε4- MCI and control groups (*P*=0.149).

For the hippocampus, significant main effects of group were also observed for FA (*P*=0.026; $F$=3.842), MD (*P*=0.009; $F$=5.007), and fiso (*P*=0.004; $F$=6.113), but not ficvf (*P*=0.445; $F$=0.643). Decreases in hippocampus FA were found in the APOE-ε4+ (ε4+: 0.133 ± 0.003; control: 0.142 ± 0.002; *P*=0.017) and APOE-ε4- (ε4-: 0.134 ± 0.003; *P*=0.039) MCI groups relative to the control group. No difference in FA was seen between APOE-ε4+ and the APOE-ε4- MCI groups (*P*=0.757). Relative to the control group, elevated hippocampal MD was found in APOE-ε4+ MCI group (APOE-ε4+: $1.027\times10^{-3}$ mm$^2$/s ± $3.0\times10^{-5}$ mm$^2$/s; control: $9.09\times10^{-4}$ mm$^2$/s ± $2.3\times10^{-5}$ mm$^2$/s; *P*=0.003). No difference was observed between hippocampal MD in APOE-ε4+ and APOE-ε4- MCI groups (APOE-ε4-: $0.967\times10^{-3}$ mm$^2$/s ± $2.9\times10^{-5}$ mm$^2$/s; *P*=0.200) or between APOE-ε4- MCI and control groups (*P*=0.093). Increased hippocampus fiso was seen in the APOE-ε4+ MCI group relative to controls (ε4+: 0.355 ± 0.020; control: 0.264 ± 0.016; *P*=0.001) but no difference was observed between APOE-ε4+ MCI and APOE-ε4- MCI groups (ε4-: 0.300 ± 0.019; *P*=0.090). No difference was seen in hippocampal fiso between APOE-ε4- MCI and control groups (*P*=0.118).

### 3.3 Tau and Susceptibility

A significant correlation between tau-PET SUVR and susceptibility was observed in the temporal lobe ($r$=0.652; *P*=0.001) and hippocampus ($r$=0.605; *P*=0.003) of the APOE-ε4+ MCI group. No association was observed between tau-PET SUVR and susceptibility in either ROI in the APOE-ε4- MCI (temporal: $r$=0.246; *P*=0.155; hippocampus: $r$=0.010; *P*=0.484) or control (temporal: $r$=0.006; *P*=0.487; hippocampus: $r$=0.109; *P*=0.286) groups. This relationship in the APOE-ε4+ MCI group, as well as their aforementioned elevated iron levels, indicate iron-related off-target binding effects in tau-PET SUVR.

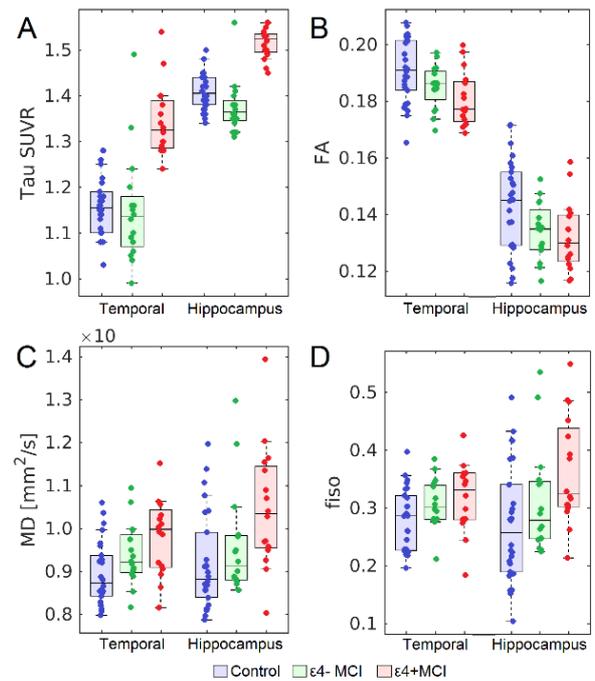

**Figure 3.** Group comparisons of tau and microstructural measures. Increases in tau SUVR (A), MD (C), and fiso (D) were seen in the temporal lobe and hippocampus ROIs of the APOE-ε4+ group relative to the control group. Decreases in FA (B) were found in the APOE-ε4+ group relative to the control group in temporal lobe and hippocampus ROIs.

### 3.4 Tau and Tissue Microstructure

Separate multiple regressions were used to examine the impact of iron on the relationship between tau-PET SUVR and tissue microstructure (MD, FA, fiso, ficvf) with (model 1) and without (model 2) controlling for susceptibility in each ROI and each clinical group. For the APOE-ε4+ MCI group, both models were significant for temporal lobe MD (model 1: $R^2$=0.391, *P*=0.006; model 2: $R^2$=0.187, *P*=0.017). Iron was a significant predictor in model 1 for MD ($\beta$=0.551, *P*=0.017), and removing it in model 2 resulted in a significant $R^2$ change ($R^2$ change=-0.225, *P*=0.017). For temporal lobe FA, only model 1 was significant (model 1: $R^2$=0.412, *P*=0.004; model 2: $R^2$=0.143, *P*=0.056) and iron was a significant predictor ($\beta$=0.570, *P*=0.007). Removing iron in model 2 resulted in a significant $R^2$ change ($R^2$ change=-0.285, *P*=0.007). For temporal lobe fiso, only model 1 was significant (model 1: $R^2$=0.364, *P*=0.008; model2: $R^2$=0.092, *P*=0.104) and iron was a significant predictor ($\beta$=0.612, *P*=0.009). Removing iron in model 2 resulted in a significant $R^2$ change ($R^2$ change=-0.291, *P*=0.009). For temporal lobe ficvf, only model 1 was significant (model 1: $R^2$=0.371, *P*=0.008; model2: $R^2$=-0.056, *P*=0.977) and iron was a significant predictor ($\beta$=0.679, *P*=0.002). Removing iron in model 2 resulted in a significant $R^2$ change ($R^2$ change=-0.437, *P*=0.002). These relationships are summarized in Figure 4.

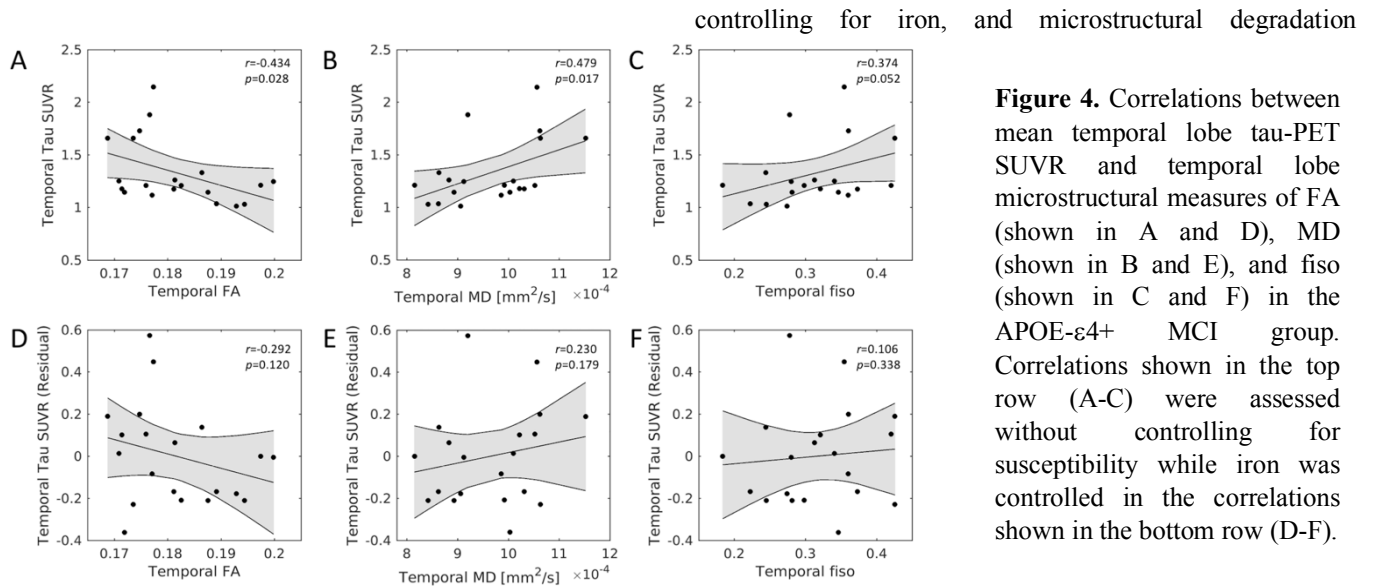

**Figure 4.** Correlations between mean temporal lobe tau-PET SUVR and temporal lobe microstructural measures of FA (shown in A and D), MD (shown in B and E), and fiso (shown in C and F) in the APOE-ε4+ MCI group. Correlations shown in the top row (A-C) were assessed without controlling for susceptibility while iron was controlled in the correlations shown in the bottom row (D-F).

Neither model was significant for hippocampal microstructure in the APOE-ε4+ MCI group (*Ps*>0.453) or for microstructure in either ROI for the APOE-ε4- and control groups (*Ps*>0.162).

### 3.3 Tau and Cognition

The relationship between hippocampal tau and cognition in the APOE-ε4+ MCI group is shown in Figure 5. Separate multiple regressions were used to examine the impact of iron on the relationship between tau-PET SUVR and cognition (ADAS delayed word recall, ADAS13) with (model 1) and without (model 2) controlling for susceptibility in each ROI and each clinical group. For ADAS delayed recall in the APOE-ε4+ MCI group, both models were significant in the hippocampus (model 1: $R^2$=0.279, *P*=0.029; model 2: $R^2$=0.318, *P*=0.007). In model 1, hippocampus tau-PET SUVR was a significant predictor ($\beta$=0.598, *P*=0.009). Removing iron in model 2 did not result in a significant change in $R^2$ ($R^2$ change=-0.003, *P*=0.800). Neither model was significant for ADAS13 in the APOE-ε4+ MCI group (*Ps*>0.150) or for either cognitive measure in the temporal lobe ROI for the APOE-ε4- and control groups (*Ps*>0.164).

### 3. Discussion

This study examines the relationship between tau deposition and both microstructural measures associated with neurodegeneration and cognition as a function of APOE-ε4 carrier status, controlling for the off-target binding effects of iron. Increased tau pathology (tau-PET SUVR), after controlling for iron, and microstructural degradation (increased MD, increased fiso, decreased FA) was found in the temporal lobe and hippocampus of APOE-ε4+ MCI participants as compared to APOE-ε4 negative MCI and control participants. Tau pathology in the hippocampus was significantly related to memory, but only in APOE-ε4+ participants. Interestingly, correlations between hippocampal tau-PET SUVR and cognitive correlations did not significantly differ when correcting for the influence of iron in the tau-PET signal. This suggests that, although a surrogate measure of iron content (susceptibility) is correlated with tau-PET SUVR in the APOE-ε4+ MCI group, the concentration of iron may not be sufficient to compromise tau-PET signal in the hippocampus.

Tissue susceptibility is associated with iron content in grey matter [39]. Several studies have reported a relationship between cortical susceptibility and tau-PET SUVR in cognitively impaired individuals.[37, 38] We found that tau-PET SUVR in the temporal lobe and hippocampus is associated with iron content in these structures of APOE-ε4+ cognitively impaired individuals. Histological[43, 44] and imaging[58] studies have reported elevated cortical iron levels in cognitively impaired individuals with the APOE-ε4 allele. No association between susceptibility and tau-PET SUVR in temporal lobe and hippocampus in the APOE-ε4- MCI and control groups. The lack of association between tau-PET SUVR and susceptibility in the temporal lobe and hippocampus of these groups may be due to low tau deposition or iron deposition in these structures. The lack of association between tau-PET SUVR and susceptibility in the control group is in agreement with an earlier study which found no association in individuals without cognitive impairment.[37]

Iron is an off-target bind to the [18]F-AV1451 radioligand.[33-36] In APOE-ε4+ subjects, this off-target binding may contribute to cortical tau-PET SUVR as the APOE-ε4 allele is linked to elevated iron levels in the brain or cerebrospinal fluid.[40, 43, 44] Using the [18]F-AV1451 radioligand, and controlling for iron-related off-target binding with tissue susceptibility, we found the APOE-ε4 allele is associated with higher tau burden in the temporal lobe and hippocampus of cognitively impaired individuals with the APOE-ε4 allele. These results agree with earlier studies in animal models[15, 16] and human stem cell models[17] that found expression of APOE-ε4 increases tau deposition. These results are comparable to earlier studies using [18]F-AV1451, which not compensate for off-target binding effects, and found greater tau burden in the temporal lobe of cognitively impaired individuals with the APOE-ε4.[59, 60] However, one study did not find temporal lobe tau deposition is influenced by the APOE-ε4 allele in individuals with mild AD.[61] This result may be related small numbers of cognitively impaired individuals with the APOE-ε4 allele or to the cognitively impaired cohort having more advanced disease.

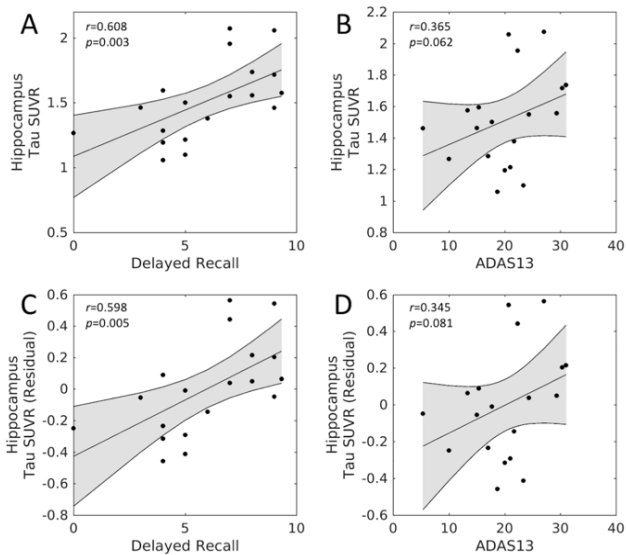

**Figure 5.** The relationship between hippocampal tau and cognition in the APOE-ε4+ MCI group. Significant correlations were seen between hippocampus tau-PET SUVR and ADAS delayed recall without controlling for iron (A) and controlling for iron (B). The relationship between hippocampus tau-PET SUVR and ADAS 13 before and after controlling for susceptibility are shown in B and D, respectively.

Prior work examining the relationship between single-compartment measures tau-PET SUVR found diffusion measures are correlated with tau-PET SUVR in the temporal lobe and hippocampus.[62, 63] We found a significant correlation between tau-PET SUVR and single-compartment diffusion measures (MD and FA) in the temporal lobe of APOE-ε4+ subjects without controlling for susceptibility, but no significant correlations remained after controlling for iron. Both tau-PET SUVR[33-36] and single compartment diffusion measures are correlated with iron content.[57] As earlier studies did not account for iron, is likely that the significant correlations observed between microstructure and tau-PET SUVR in earlier studies[62, 63] are due to the influence of iron.

Hyperphosphorylated tau is associated with AD-related cognitive impairment[64] and this association has motivated the idea that tau initiates neurodegeneration.[9] Imaging studies have observed tau burden is closely related to cortical thinning in individuals with MCI or AD[65, 66] and tau deposition has been found to precede cortical thinning.[67] In the APOE-ε4+ MCI group, we found higher tau burden and changes in microstructural measures related to neurodegeneration (higher MD, higher fiso, and lower FA) in the temporal lobe and hippocampus relative to the APOE-ε4- MCI and control groups. These results are in agreement with earlier studies since cortical thinning should increase the unrestricted diffusion compartment (i.e. fiso) in the NODDI model thereby increasing MD and decreasing FA.

The APOE-ε4 allele is linked to higher cognitive impairment[11-13] in individuals with AD or MCI. In agreement with these studies,[11-13] we found MCI participants with the APOE-ε4 allele to have greater cognitive impairment as compared to MCI or control participants without the APOE-ε4 allele. As detailed above, this impairment may be related to tau pathology [9, 28] since hippocampal tau-PET SUVR was found to be strongly related to ADAS delayed word recall in the APOE-ε4+ MCI group with higher memory impairment correlated with higher hippocampal tau burden. Taken together, these results agree with hypotheses that hippocampal tau accumulation is associated with cognitive decline.[9] However, given the small sample size, results linking APOE-ε4 status to tau accumulation and cognitive decline should be interpreted with caution.

This study has several caveats. First, while QSM is sensitive to iron[39] and is also sensitive to myelin.[68] Grey matter regions in the hippocampus and temporal lobe with low myelin content were investigated in this study. While white matter regions were excluded in our analysis of cortical iron and tau deposition, partial volume effects may include white matter and bias cortical susceptibility measurements. Second, iron deposition has been hypothesized to accelerate tau hyperphosphorylation.[69] Off-target binding of the [18]F-AV1451 radioligand to iron impedes investigation of the relationship between iron deposition and tau hyperphosphorylation.

In this work, tissue susceptibility was used to account for iron-related off-target binding effects with the [18]F-AV1451 radioligand in the hippocampus and temporal lobe in cognitively normal and impaired individuals. In the APOE-ε4+ group, we found a correlation between uptake of the [18]F-AV1451 radioligand (tau-PET SUVR) and tissue susceptibility in the hippocampus and temporal lobe. Individuals in the APOE-ε4+ MCI group were found to have higher tau burden in the hippocampus and temporal lobe, after accounting for tissue susceptibility, than individuals in the control or APOE-ε4- MCI groups. We found off-target binding to have no effect on correlations between cognitive measures and hippocampal tau-PET SUVR in the APOE-ε4+ MCI group. Increases in MD and fiso were also observed in the hippocampus and temporal lobe of APOE-ε4+ MCI group relative to control and APOE-ε4- MCI groups. Controlling for tissue susceptibility was found to influence correlations between tau-PET SUVR and diffusion metrics and the change in this interaction may be due to the influence of iron on diffusivity. Taken together, these results suggest that iron does not need to be accounted for in group comparisons of tau-PET SUVR or correlations between cognitive measures and tau-PET SUVR in the hippocampus and temporal lobe but should be accounted for when correlating with diffusion metrics derived from a single-compartment model.

## Acknowledgements

This work was supported by 1K23NS105944-01A1 (Huddleston) from the National Institutes of Health/National Institute of Neurological Diseases and Stroke, and the American Parkinson's Disease Foundation Center for Advanced Research at Emory University (Huddleston).

Data used in preparation of this article were obtained from the Alzheimer's Disease Neuroimaging Initiative (ADNI) database (adni.loni.usc.edu). As such, the investigators within the ADNI contributed to the design and implementation of ADNI and/or provided data but did not participate in analysis or writing of this report. A complete listing of ADNI investigators can be found at: http://adni.loni.usc.edu/wp-content/uploads/how_to_apply/ADNI_Acknowledgement_List.pdf